\documentstyle[preprint,aps]{revtex}
\begin{document}
\title{Quantum-Classical Phase Transition of Escape Rate in Biaxial Spin Particles}
\author{Y.-B. Zhang$^{1,2}$\footnote{email: ybzhang@physik.uni-kl.de}, J.-Q. Liang$^{1,2}$, H. J. W. M\"{u}ller-Kirsten$^{2}$\footnote{email: mueller1@physik.uni-kl.de}, S.-P. Kou$^{1,3}$, X.-B. Wang$^4$, F.-C. Pu$^{3,5}$}
\address{1. Department of Physics and Institute of Theoretical Physics, Shanxi University, Taiyuan, Shanxi 030006, China\\
2. Department of Physics, University of Kaiserslautern, D-67653 Kaiserslautern, Germany\\
3. Institute of Physics and Center for Condensed Matter Physics, Chinese Academy of Sciences, Beijing 100080, China\\
4. Center for Advanced Study and Department of Physics, Tsinghua University, Beijing 100084, China\\
5. Department of Physics, Guangzhou Normal College, Guangzhou 510400, China}

\maketitle

\begin{abstract}
The escape rates of the biaxial single domain spin particles with and without an applied magnetic field are investigated. Using the strict potential field description of spin systems developed by Ulyanov and Zaslavskii we obtain new effective Hamiltonians which are considered to be in exact spin-coordinate correspondence unlike the well studied effective Hamiltonians with the approximate correspondence. The sharp first-order transition is found in both cases. The phase diagram of the transitions depending on the anisotropy constant and the external field is also given.
\\
PACS numbers: 75.45.+j,75.50.Tt
\end{abstract}

\section{Introduction}
\label{sec:I}

The decay rate of metastable states or transition rate between degenerate vacua is dominated at high temperatures by thermal activation, whereas at temperatures close to zero, quantum tunneling is relevant. At some critical temperature the transition from the classical to the quantum-dominated regime occurs. The transition can be first-order, with a discontinuous first derivative of the escape rate, or smooth with only a jump of the second derivative in which case it is known as of second-order.

Based upon the functional-integral approach Affleck\cite{Affleck} and Larkin and Ovchinnikov \cite{Larkin} demonstrated with certain assumptions for the shape of the potential barrier that a second-order phase transition from the thermal to the quantum regime takes place at a critical temperature $T_0=1/\beta_0$, where $\beta_0$ is the period of small oscillations near the bottom of the inverted potential well. Chudnovsky, however, showed that the situation is not generic and that the crossover from the thermal to the quantum regime can quite generally be the first-order transition\cite{Chudnovsky1} that takes place at $T_c>T_0$ for the case in which the period versus energy curve possesses a minimum. Shortly after the observation of Chudnovsky the sharp first-order transitions were found theoretically in spin tunneling for two systems. One of these is a ferromagnetic bistable large-spin particle\cite{Chudnovsky2,Garanin1} described by the Hamiltonian ${\hat H}=-D{\hat S}_z^2-B_x{\hat S}_x$ which is believed to be a good approximation for the molecular magnet $Mn_{12}Ac$ of spin $S=10$, and the other is a biaxial anisotropic model, whose effective mass was shown to be position-dependent\cite{Liang1}. It was the external field $B_x$ (in first model) and the anisotropic constant ratio $\lambda$ (in second model) that effect the phase transition of the crossover. The same models with the magnetic field applied along alternative axes have also been studied, and the corresponding phase diagrams have been given\cite{Lee1,Garanin2}. A sufficient criterion for the first-order transition in the context of tunneling can be obtained by studying the Euclidean time period in the neighbourhood of the sphaleron configuration at the peak of the potential barrier\cite{Gorokhov,Muller-Kirsten}. In the present paper we incorporate the two parameters in a single spin tunneling model in order to investigate the dependence of the phase transition on these.

The phenomenon of spin tunneling has attracted considerable attention not only in view of the possible experimental test of the tunneling effect for mesoscopic single domain particles -  in which case it is known as macroscopic quantum tunneling - but also because the spin system with an applied field provides various potential shapes and therefore serves as a testing ground for theories of instanton induced transitions. The key procedure in dealing with spin tunneling is to convert the discrete spin system into a continuous one by a spin-coordinate correspondence. There are various spin variable techniques which result in effective Hamiltonians. It is a long standing question whether the different effective Hamiltonians for a given quantum spin system lead to the same result \cite{Ulyanov,Hemmen}. Following Ulyanov and Zaslavskii we have obtained a new effective Hamiltonian for the spin particle with biaxial anisotropy in addition to the one of a sine-Gordon potential with position dependent mass already known in the literature\cite{Liang1,Lee1,Enz,Liang2}.

The paper is organized as follows. In Sec. II we give a brief review of the general theory of phase transitions of escape rates. Using the effective method of Ref. \cite{Ulyanov} we then derive in Sec. III an alternative effective Hamiltonian for the ferromagnetic particle with a biaxial anisotropy without an applied magnetic field. It is shown that the sharp first-order transition from the classical to the quantum regime indeed exists in agreement with the observation in our previous paper \cite{Liang1}. In Sec. IV we then investigate the spin tunneling and phase transition with an external field applied along the easy axis which breaks the symmetry and makes one of the degenerate vacua metastable. Our conclusions and discussions are given in Sec. V.

\section{The Criterion for the Sharp First-Order Phase Transition of the Escape Rate}
\label{sec:II}

At temperature $T$ the escape rate of a particle through a potential barrier can be obtained by taking the ensemble average of the tunneling probability $\Gamma_t(E)$, i. e.
\begin{equation}
  \Gamma(T)=\int \Gamma_t(E)e^{-\frac ET} dE
\end{equation}
where the tunneling probability at a given energy $E$ is defined by
\begin{equation}
  \Gamma_t(E)=Ae^{-W(E)}  \label{gt}
\end{equation}
and
\begin{equation}
  W(E)=2\int_{\phi_i(E)}^{\phi_f(E)}d \phi \sqrt{2m(\phi)[V(\phi)-E]}
\end{equation}
is evaluated from the periodic pseudoparticle (instanton or bounce) trajectories $\phi_c$ between turning points $\phi_i$ and $\phi_f$. The pseudoparticle trajectory $\phi_c$ minimizes the Euclidean action at the given energy $E$ above the metastable minimum such that $\delta S(\phi_c)=0$ with periodic boundary condition $\phi_c(0)=\phi_c(\beta)$. The Euclidean action $S_E$ and Lagrangian ${\cal L}_E$ are
\begin{equation}
  S_E=\int d\tau {\cal L}_E=W+\beta E, \quad
  {\cal L}_E=\frac 12 m(\phi) \dot\phi^2+V(\phi)
\end{equation}
respectively. Here $\dot\phi \equiv d\phi/d\tau$ and $\tau=it$ denotes Euclidean time. In general the mass $m(\phi)$ could be position dependent in the context of spin tunneling. The time period $\beta(E)$ is related to temperature $T$ by $\beta(E)=\frac 1T$, as usual. The prefactor $A$ in Eq. (\ref{gt}) results from a Gaussian functional integration over small fluctuations around the pseudoparticle trajectory $\phi_c$. In the semiclassical approximation the escape rate at temperature $T$ is dominated by
\begin{equation}
  \Gamma(T)\sim e^{-S_{min}(T)},
\end{equation}
where $S_{min}(T)$ is the minimum effective Euclidean action which is chosen as the smallest value of $S_0$ and $S(T) \equiv S_E$. Here, $S_0$ is the thermodynamic action defined by
\begin{equation}
  S_0=\beta E_0 \label{S0}
\end{equation}
with $E_0$ being the barrier height for the pseudoparticle to tunnel through. Generally speaking, at $E=0$ (the bottom of the initial well) the Euler-Lagrange equation leads to the vacuum instanton or bounce solution. When $0<E<E_0$ (between the bottom and the top of the barrier) the trajectory $\phi(\tau)$ shows periodic motion in the barrier region of the potential $V(\phi)$ which is forbidden for the classical particle. The period of oscillation as a function of energy $E$ is given by the following integral
\begin{equation}
  \beta (E)=\int_{\phi_i(E)}^{\phi_f(E)} d \phi \frac {\sqrt {2 m(\phi)}}{\sqrt {V(\phi)-E}}.
\end{equation}

There are two independent criteria for the existence of first-order transitions between the classical and quantum regimes. The non-monotonic behavior of the oscillation period as a function of energy, i.e. the existence of a minimum in the $\beta \sim E$ curve, was once proposed as a condition\cite{Chudnovsky1} for first-order phase transitions in quantum mechanical tunneling. The corresponding dependence of the action on temperature leads to an abrupt change at some critical temperature, at which the first derivative of $S_{min}(\beta)$ is discontinuous, indicating that the crossover from the thermal to the quantum regime is the first-order transition in temperature.

More generally, for a massive particle with position coordinate $q$, it has been shown \cite{Muller-Kirsten} that the existence of a first-order transition leads to the condition
\begin{equation}
  \left[ V'''(q_s)(g_1+\frac{g_2}2)+\frac 18V''''(q_s)+M'(q_s)\omega^2g_2+M'(q_s)(g_1+\frac{g_2}2)+\frac 14M''(q_s)\omega^2\right]_{\omega_0}<0 \label{cr}
\end{equation}
where
\begin{eqnarray}
  g_1(\omega)=-\frac{\omega^2M'(q_s)+V'''(q_s)}{4V''(q_s)}, \nonumber \\
  g_2(\omega)=-\frac{2M'(q_s)+V'''(q_s)}{4[4M(q_s)\omega_0^2+V''(q_s)]}.
\end{eqnarray}
and
\begin{equation}
  \omega_0^2=-\omega_s^2=\frac{V''(q_s)}{M(q_s)} \label{omega0}
\end{equation}
and $M$ is the position dependent mass. The subscript $s$ stands for the coordinate of the bottom of the well of the inverted potential, i.e., the coordinate of the sphaleron. This criterion has been applied to various models studied earlier and the results coincide with previous ones.

\section{Effective Hamiltonian of the Ferromagnetic Particle}
\label{sec:III}

The model we consider here is that of a nanospin particle which is assumed to have a biaxial anisotropy with XOY easy plane and the easy X-axis in the XY-plane, and is  described by the Hamiltonian
\begin{equation}
  {\hat H}=K_1\hat{S}_z^2+K_2\hat{S}_y^2,\qquad K_1>K_2>0 \label{h1}
\end{equation}
which has been extensively studied in the context of tunneling from various aspects such as ground state tunneling\cite{Enz,Chudnovsky3}, tunneling at finite energy, namely, with the periodic instanton \cite{Liang2}, and topological quenching of tunneling \cite{Loss,Liang4}. Most recently it was shown that this model possesses a first-order phase transition from the thermal to the quantum regime \cite{Liang1}. In all these investigations the quantum spin system of Eq. (\ref{h1}) is converted into a potential problem by using the conventional spin coherent state technique with approximate spin-coordinate correspondence (see Appendix). The effective potential is of the sine-Gordon type.

In the present investigation we reexamine the quantum spin system in terms of a new method developed by Ulyanov and Zaslavskii\cite{Ulyanov}. The spin operator representation in differential form on the basis of spin coherent states is given by the relations
\begin{equation}
  \hat{S}_z=-i\frac d{d\varphi },\qquad \hat{S}_x=S\cos \varphi -\sin \varphi \frac d{d\varphi },\qquad \hat{S}_y=S\sin \varphi +\cos \varphi \frac d{ d\varphi } \label{sc}
\end{equation}
The eigenvalue equation
\begin{equation}
  {\hat H}\left| \Phi \right\rangle =E\left| \Phi \right\rangle 
\end{equation}
then becomes a second-order differential equation, i.e.
\begin{eqnarray*}
  \left( K_1-K_2\sin ^2\phi \right) \frac{d^2\Phi }{d\phi ^2}+K_2\left( S-%
  \frac 12\right) \sin 2\phi \frac{d\Phi }{d\phi }+\left( E-K_2S^2\cos
  ^2\phi -K_2S\sin ^2\phi \right) \Phi =0 \nonumber
\end{eqnarray*}
where we have shifted the azimuthal angle by $\frac\pi2$ for convenience, $\varphi =\phi +\frac \pi 2$. Following Ref. \cite{Ulyanov}, we use the transformation
\begin{eqnarray}
  \Psi &=&\Phi (\phi )(K_1-K_2\sin ^2\phi )^{-\frac S2}, \nonumber \\
  x &=&\int_0^\phi \frac{d\phi ^{\prime }}{\sqrt{1-\lambda \sin ^2\phi
  ^{\prime }}}=F(\phi ,\tilde{k}),\quad \tilde{k}^2=\lambda=\frac{K_2}{K_1},\quad \sin \phi ={\rm sn}x
\end{eqnarray}
The eigenvalue equation is then transformed to the following effective potential form
\begin{equation}
  -K_1\frac{d^2\Psi }{dx^2}+K_2S(S+1)\frac{{\rm cn}^2x}{{\rm dn}^2x}\Psi =E\Psi 
\end{equation}
where ${\rm sn}x$, ${\rm cn}x$ and ${\rm dn}x$ denote Jacobian elliptic functions. The effective Hamiltonian is seen to be
\begin{equation}
  {\hat H}=\frac{p^2}{2m}+U(x),\qquad m=\frac 1{2K_1},\qquad U(x)=K_2S(S+1) {\rm cd}^2x. 
\end{equation}
with ${\rm cd}x={\rm cn}x/{\rm dn}x$. We remark here that this derivation, unlike that in previous investigations, is exact and without a large $s$ limiting procedure. We also emphasize that this Jacobian elliptic potential is of interest on its own and has not been investigated before in the context of instanton considerations. The periodic instanton solution leads to an integral with finite energy and is obtained as
\begin{equation}
  x_p={\rm sn}^{-1} [ k {\rm sn}(\omega \tau),\tilde{k} ]
\end{equation}
The trajectory of one instanton as half of the periodic bounce is shown in Fig. 1 with added instanton$-$anti-instanton pair. We choose $s=\sqrt{1000}$ and $K_1=1$(thus $K_2=\lambda$) in the diagrams. The period of this periodic configuration is seen to be
\begin{equation}
  \beta(E)=\frac {4{\cal K}}{\omega(E)} =\frac{2}{\sqrt{K_1}}\frac{1}{\sqrt{K_2s^2-E\lambda}} {\cal K}(k)
\end{equation}
where ${\cal K}(k)$ is the complete elliptic integral of the first kind, and
\begin{eqnarray}
  k&=&\sqrt{\frac{n^2-1}{n^2-\lambda}}, \quad n^2=\frac{K_2s^2}{E},\nonumber \\
  \omega&=&\omega_0 \sqrt{1-\frac{\lambda}{n^2}}, \quad \omega_0^2=4K_1K_2s^2.
\end{eqnarray}
The action calculated along the above trajectory is $S=W+\beta E$ with
\begin{equation}
  W =\frac \omega {\lambda K_1}\left[ K(k)-(1-\lambda k^2)\Pi (\lambda k^2,k)\right]
\end{equation}
Here we are particularly interested in the phase transition. To this end we show in Fig. 2 the shape of the potential for various values of $\lambda$. The peak of the barrier becomes flatter and flatter as $\lambda$ increases. The curve of $\beta$ versus $E$ is given in Fig. 3 and demonstrates the obvious first-order phase transition for $\lambda>1/2$. Fig. 4 shows the action as a function of temperature. 

Next we apply the criterion for the first-order phase transition to the model above. The sphaleron is located at $x_s=0$. Computing the corresponding quantities at the sphaleron position, i.e.
\begin{eqnarray}
  V[x_s]=K_2s^2, \quad V'[x_s]=0, \quad V''[x_s]=-2 K_2s^2(1-\lambda),\nonumber \\
  V'''[x_s]=0, \quad V''''[x_s]=8K_2s^2(1-\lambda)(1-2\lambda)
\end{eqnarray}
Eq. (\ref{cr}) becomes
\begin{equation}
  \frac 18V''''(x_s)=K_2s^2(1-\lambda)(1-2\lambda)<0
\end{equation}
and we regain the critical value of $\lambda_c=\frac 12$ at which the first-order transition sets in. We see the new effective Hamiltonian with exact spin-coordinate correspondence leads to the same results as those of ref. \cite{Liang1}.

However, the physical interpretation for the sharp first-order phase transition is now different. In the present case the effective mass is constant and the sharp transition from quantum to classical behavior results from a flattening of the peak of the barrier. In Ref. \cite{Liang1} the first-order transition resulted from the position dependence of the mass which makes the latter heavier at the top of the barrier.

\section{The phase transition with an applied field along the easy axis}
\label{sec:IV}

The Hamiltonian with an applied magnetic field $h$ along the easy $X$-axis is given by
\begin{equation}
  {\hat H}=K_1\hat{S}_z^2+K_2\hat{S}_y^2-g \mu_B h \hat{S}_z, \label{hh}
\end{equation}
where $\mu_B$ is the Bohr magneton, and $g$ is the spin $g$-factor which is taken to be $2$ here. The anisotropy energy associated with this Hamiltonian has two minima: the one on the $+X$-axis which is a metastable state and the other on the $-X$-axis. Between these two energy minima there exists an energy barrier, and the spin escapes from the metastable state either by crossing over or by tunneling through the barrier.

Following the same procedure as in the previous section, the Hamiltonian of Eq. (\ref{hh}) can be mapped onto a point particle problem with effective Hamiltonian
\begin{equation}
  {\hat H}=\frac {p^2}{2m}+V(x) \label{he}
\end{equation}
where
\begin{equation}
  m=\frac 1{2K_1}, \quad V(x)=K_2s^2(1+\frac {\alpha^2 \lambda}{4}) {\rm sn}^2(x,\tilde{k})+K_2s^2\alpha \left({\rm cn} (x,\tilde{k}) {\rm dn} (x,\tilde{k})-1\right)
\end{equation}
where the metastable minimum of the potential has been shifted to $x=0$ for convenience. Now the effective mass is a constant. The potential is more like that of an inverted double-well potential with the sphaleron position $x_s={\rm cd}^{-1}(\frac \alpha 2,{\tilde k})$ and barrier height
\begin{equation}
  E_0=K_2s^2(1+\frac {\alpha^2 \lambda}{4}) \frac {1-\frac{\alpha^2}4}{1-\frac{\lambda \alpha^2}4}+K_2s^2\alpha \frac {(1+\frac{\alpha \lambda}2)(\frac \alpha 2-1)}{1-\frac{\lambda \alpha^2}4}
\end{equation}
Under the barrier a bounce configuration exists as shown in Fig. 5. We redraw the period diagram as a function of energy for the same parameter ($\lambda=0.9$, $\alpha=1$) in Fig. 6 and obtain the first-order transition from thermal activation to quantum tunneling as shown in Fig. 7.

Applying the phase transition criterion to this model, we obtain
\begin{eqnarray}
  V'(x_s)&=&0,\quad V''(x_s)=-2K_2s^2(1-\frac {\alpha^2}{4})(1-\lambda),\nonumber\\
  V'''(x_s)&=&-3K_2s^2\alpha(1-\lambda)^2\sqrt{\frac{1-\frac{\alpha^2}4}{1-\frac{\alpha^2 \lambda}4}},\nonumber \\
 V''''(x_s)&=&\frac {2K_2s^2(1-\lambda)}{4-\alpha^2 \lambda} \left((\lambda^2-2\lambda)\alpha^4-(7\lambda^2-22\lambda+7)\alpha^2-32\lambda+16\right)
\end{eqnarray}
and the frequency $\omega_0^2$ is
\begin{equation}
  \omega_0^2=\omega_s^2=-\frac{V''(x_s)}{m}=-2K_1V''(x_s)
\end{equation}
The expressions for $g_1$ and $g_2$ are now found to be
\begin{eqnarray}
  g_1&=&-\frac{\frac 38 \alpha (1-\lambda)}{\sqrt{(1-\frac{\alpha^2}4)(1-\frac{\alpha^2 \lambda}4)}} \nonumber \\
  g_2&=&\frac{\frac 18\alpha(1-\lambda)}{\sqrt{(1-\frac{\alpha^2}4)(1-\frac{\alpha^2 \lambda}4)}}
\end{eqnarray}
The critical line of the two parameters for the first order transition requires
\begin{equation}
  \lambda>\frac{2(2+\alpha^2)}{8+\alpha^2}
\end{equation}
The corresponding phase diagram is shown in Fig. 8. From the diagram we observe several interesting features. First, the classical-quantum phase transition shows both the first-order (region I) and the second-order (region II) transition domains. We see that there is only a second-order transition for $\lambda<0.5$. For materials with $\lambda$ larger than $0.5$ we can see that the order of the phase transition changes from first to second as $\alpha$ increases and the phase boundary changes with $\lambda$ up to $1$. 

An alternative effective Hamiltonian\cite{Liang3,Park} with the conventional application of the spin coherent state technique as that for the biaxial anisotropy spin particle without the applied magnetic field \cite{Liang1} has also been investigated. However, the effective Hamiltonian with approximate spin-coordinate correspondence gives rise to a result \cite{Park} for the phase transition which differs from that of the Hamiltonian of Eq. (\ref{he}) with exact spin-coordinate correspondence. The center of the position dependent mass in the Hamiltonian with approximate spin-coordinate correspondence in Refs.\cite{Liang3,Park} does not coincide with the position of the pseudoparticle, namely, the bounce here. We, therefore, believe that the instanton concept fails in this case. A detailed analysis will be reported elsewhere.

\section{Conclusion}
\label{sec:V}

In summary, we investigated the biaxial spin model with applied magnetic field along the easy $X$-axis with the effective potential methods developed in Ref. \cite{Ulyanov}. The resulting periodic instanton solutions with potential in terms of Jacobian elliptic functions and the corresponding phase transitions have been studied for the first time. For the quantum spin particles with anisotropic constant $\lambda>\frac 12$ and above the critical value of magnetic field $\alpha$, the energy dependence of the oscillation period shows the obvious first-order transition from thermal activation to thermally assisted quantum tunneling. Applying the criterion for the first-order transition, we obtained the phase diagram which exhibits two domains separated by a critical line, indicating the first-order and second-order transitions respectively.

\begin{center}
  \bf{Appendix}
\end{center}

In the conventional application of the spin coherent state technique, two canonical variables, $\phi$ and $p=s \cos \theta$ are adopted with the usual quantization
\begin{equation}
  [\phi,p]=i \eqnum{A1} \label{qr}
\end{equation}
We show in the following that the spin-coordinate correspondence is only approximate up to order $O(s^{-3})$.

From the relation between the spin operators and the polar coordinate angles
\begin{equation}
  S_x=s\sin\theta\cos\phi,\quad S_y=s\sin\theta\sin\phi, \quad S_z=s\cos\theta \eqnum{A2}
\end{equation}
the usual commutation relation of spin operators reads
\begin{equation}
  [S_x,S_y]=s^2[\sin\theta\cos\phi,\sin\theta\sin\phi]=s^2\sin\theta[\cos\phi,\sin\theta]\sin\phi+s^2\sin\theta[\sin\theta,\sin\phi]\cos\phi \eqnum{A3} \label{a12}
\end{equation}
Using Eq. (\ref{qr}), one can prove the following relation
\begin{equation}
  [\sin\theta,\cos\phi] = A_+\cos\phi+iA_-\sin\phi,\quad
  [\sin\theta,\sin\phi] = A_+\sin\phi-iA_-\cos\phi \eqnum{A4} \label{a4}
\end{equation}
with
\begin{eqnarray*}
  A_+=\frac 12 \left(\sqrt{1-(\cos\theta+\alpha)^2}+\sqrt{1-(\cos\theta-\alpha)^2} \right),\\
  A_-=\frac 12 \left(\sqrt{1-(\cos\theta+\alpha)^2}-\sqrt{1-(\cos\theta-\alpha)^2} \right)
\end{eqnarray*}
where $\alpha=1/s$. Substituting (\ref{a4}) into Eq. (\ref{a12}), one has
\begin{equation}
  [S_x,S_y]=s^2(-i)\sin\theta A_-=i\cos\theta s^2\alpha+O(\alpha^3) \eqnum{A5}
\end{equation}
i.e.
\begin{equation}
  [S_x,S_y]=i S_z+O(s^{-3}) \eqnum{A6}
\end{equation}
which implies that the usual commutation relation holds only in the large spin limit.

{\bf Acknowledgment:} This work was supported by the National Natural Science Foundation of China under Grant Nos. 19677101 and 19775033. J.-Q. L. also acknowledges support by a DAAD-K.C.Wong Fellowship.

\newpage
\begin{center}
  {\bf Figure Captions:}
\end{center}
\noindent
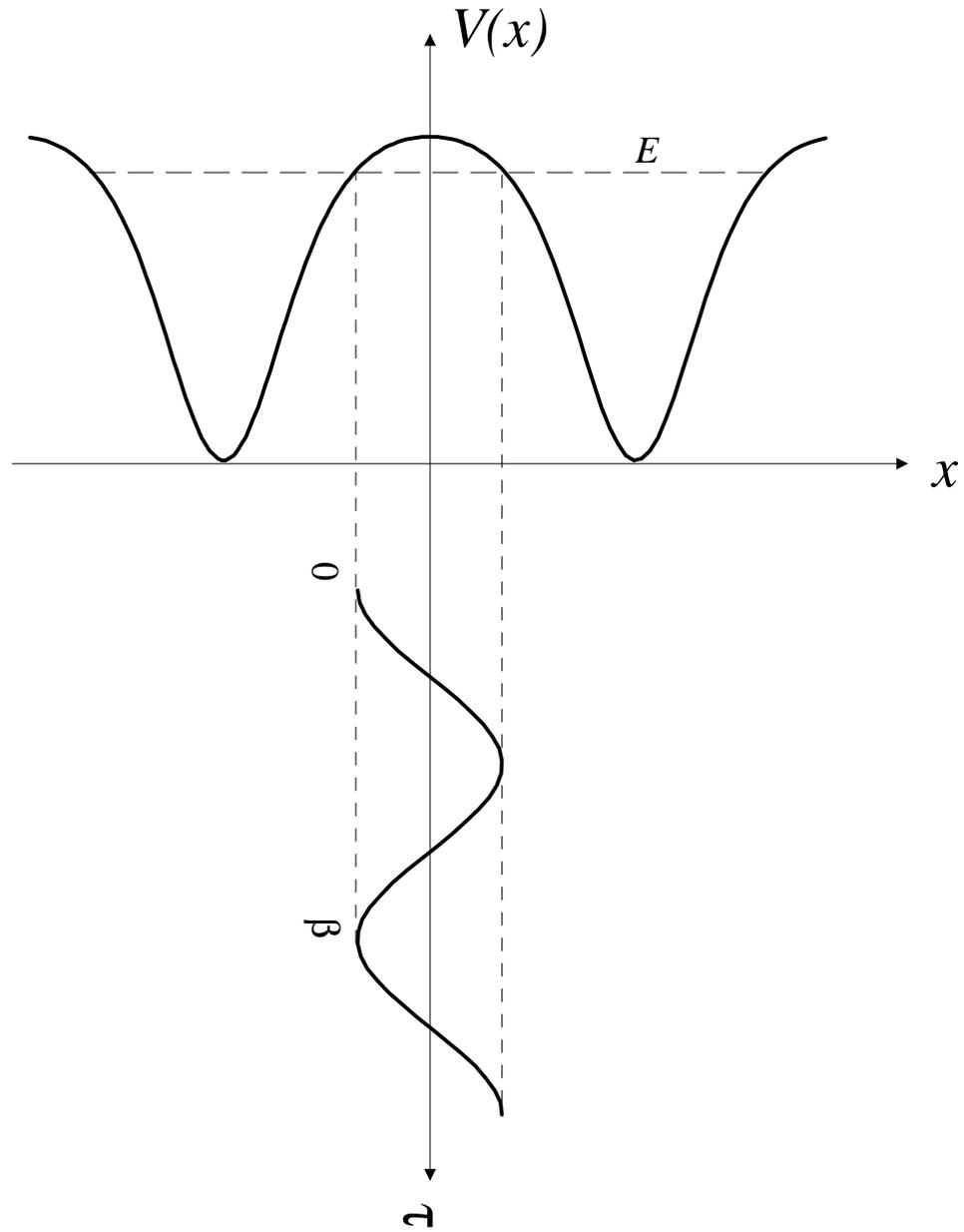
Fig. 1: The effective potential and the corresponding periodic instanton configurations.\\
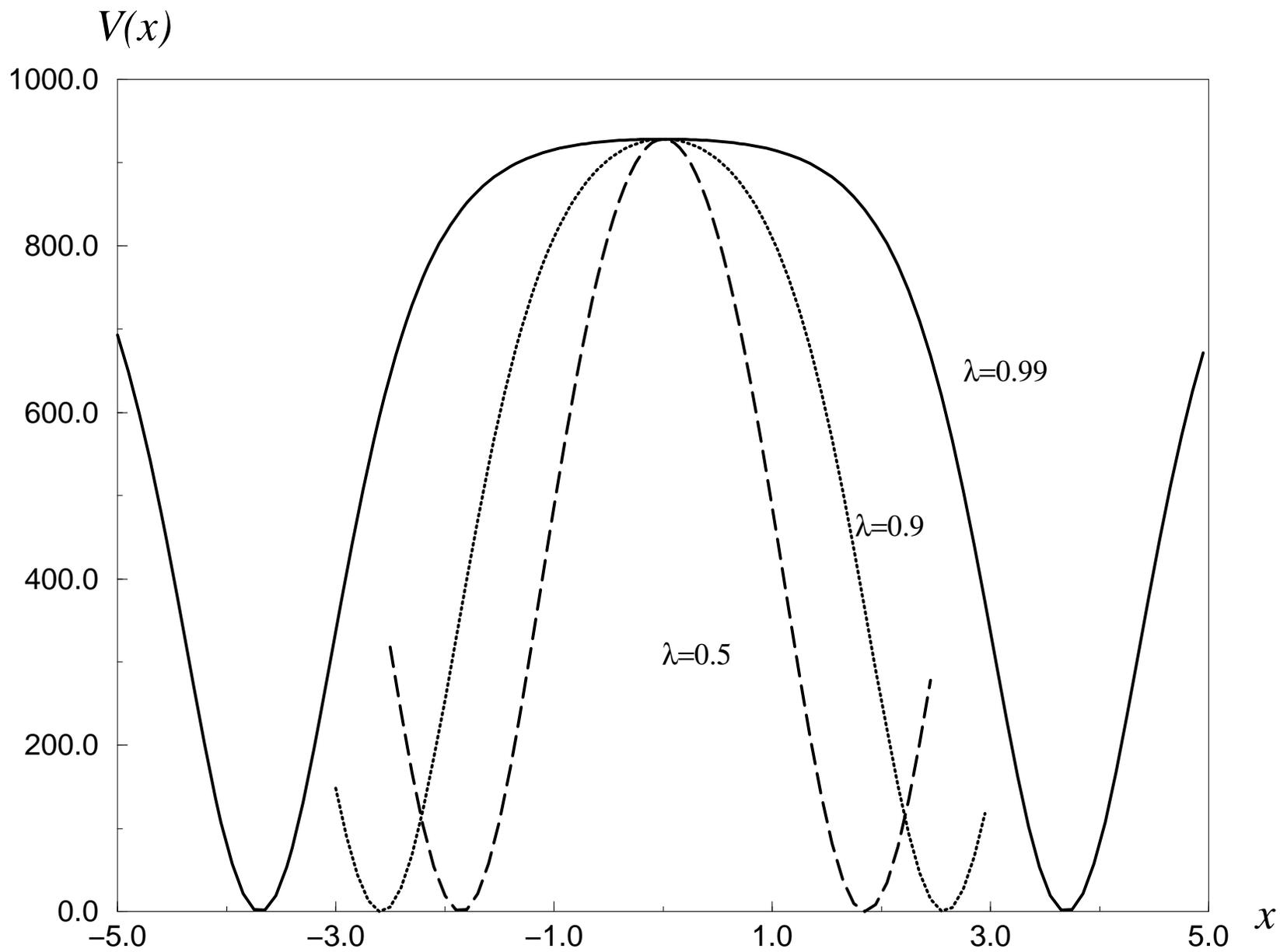
Fig. 2: The potentials for different values of $\lambda$.\\
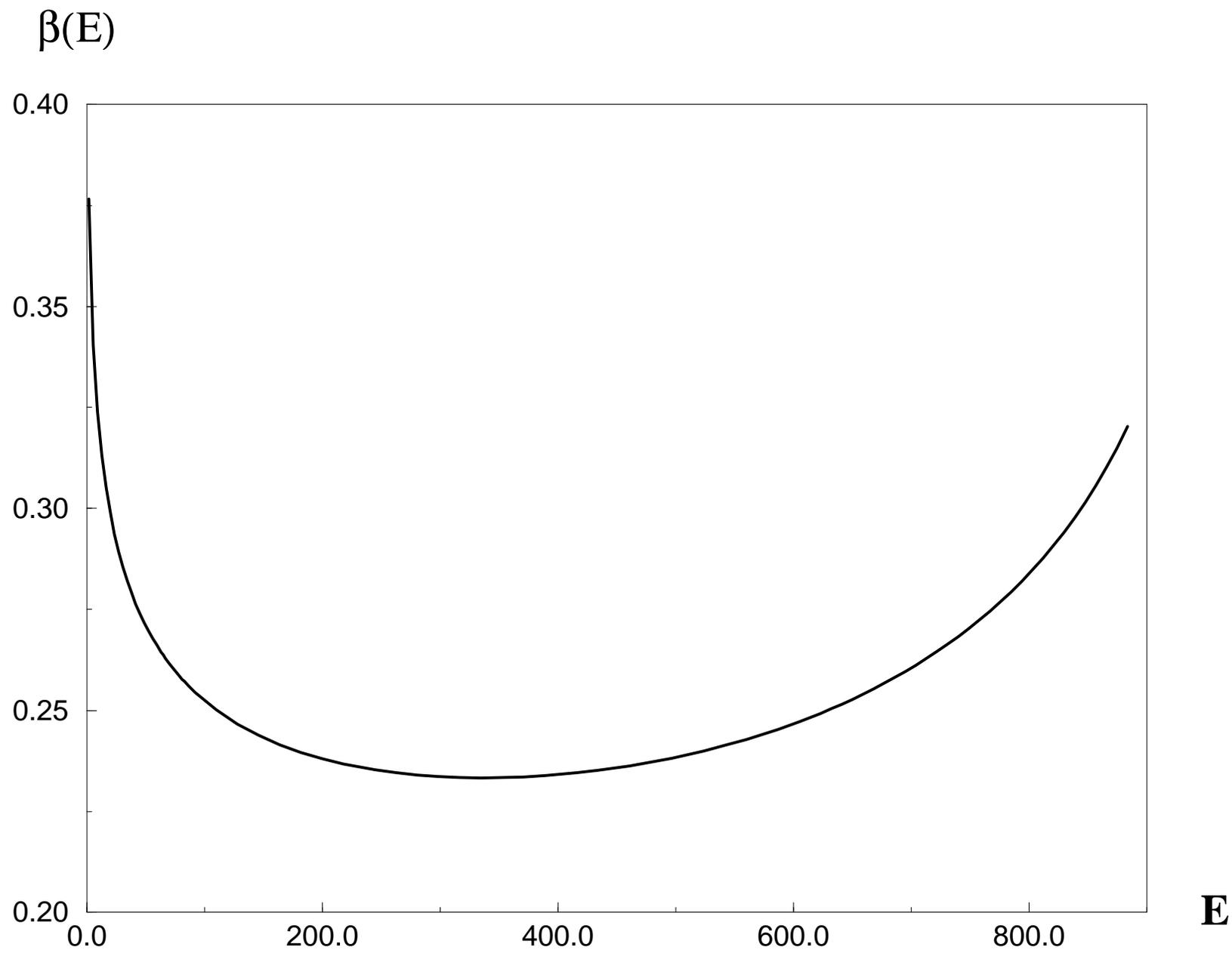
Fig. 3: The oscillation period as a function of energy at $\lambda=0.9$.\\
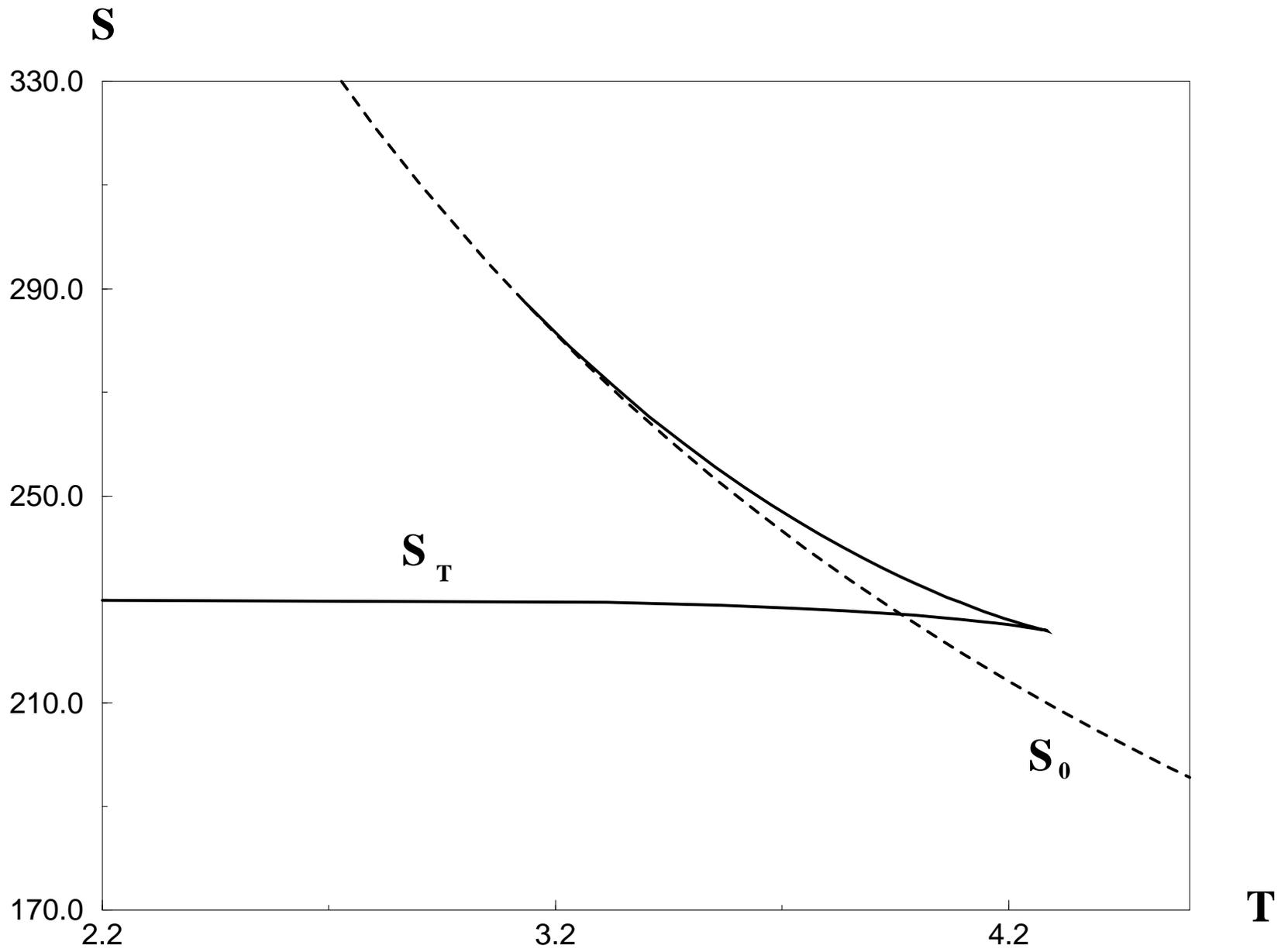
Fig. 4: The action as a function of temperature indicating the first-order transition at $\lambda=0.9$.\\
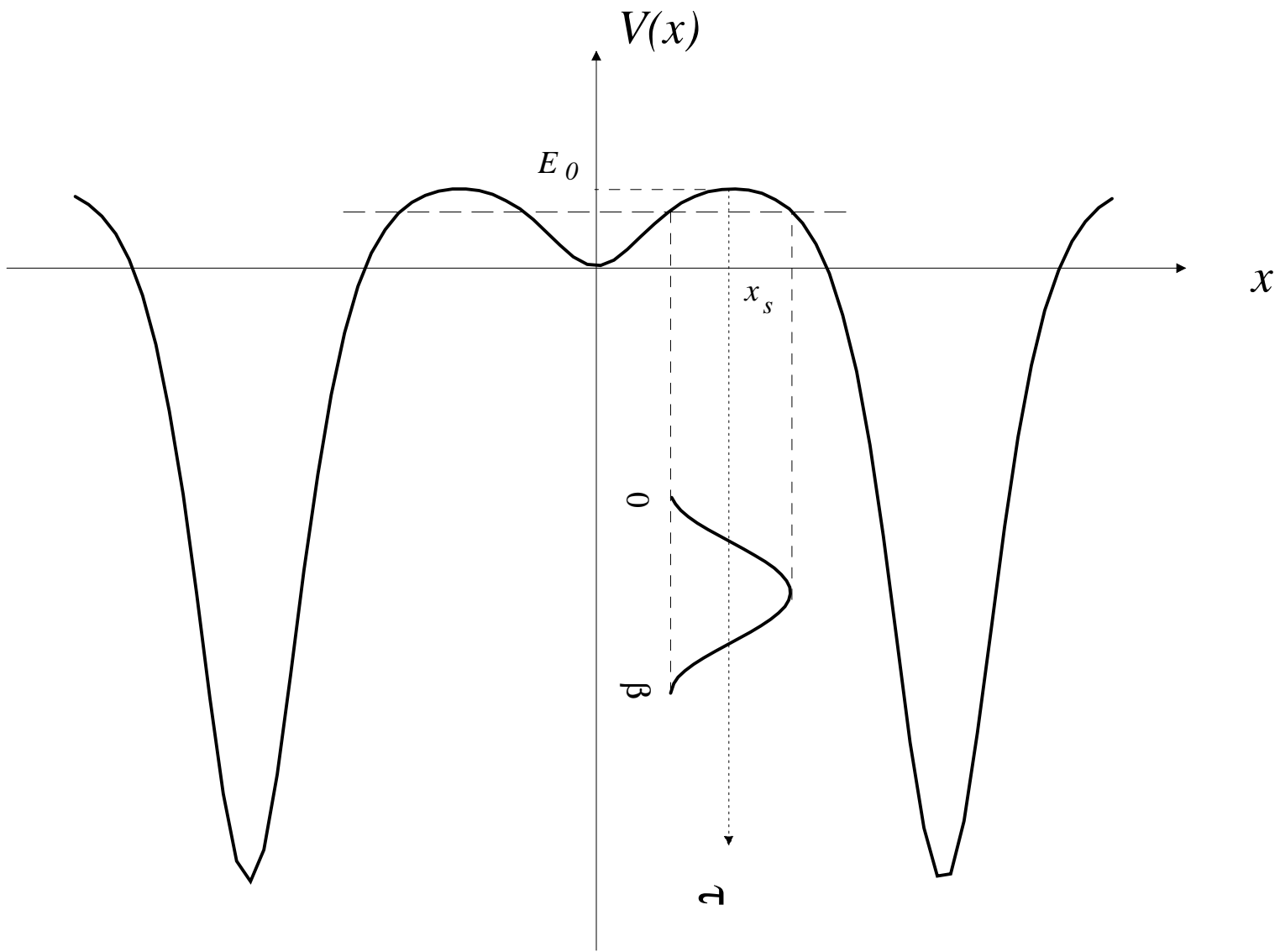
Fig. 5: The effective potential in the case with external magnetic field.\\
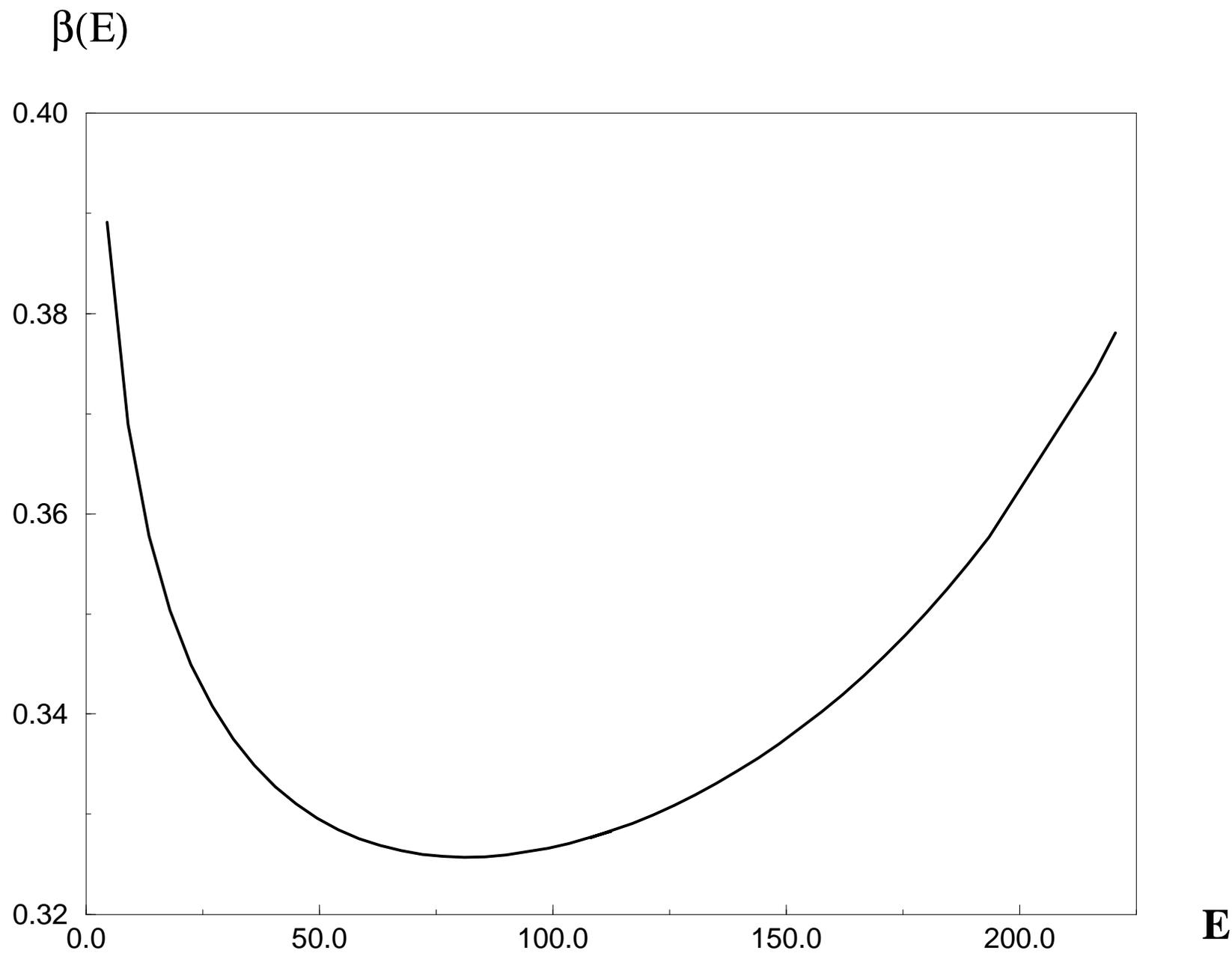
Fig. 6: The oscillation period as a function of energy at $\lambda=0.9$ and $\alpha=1$.\\
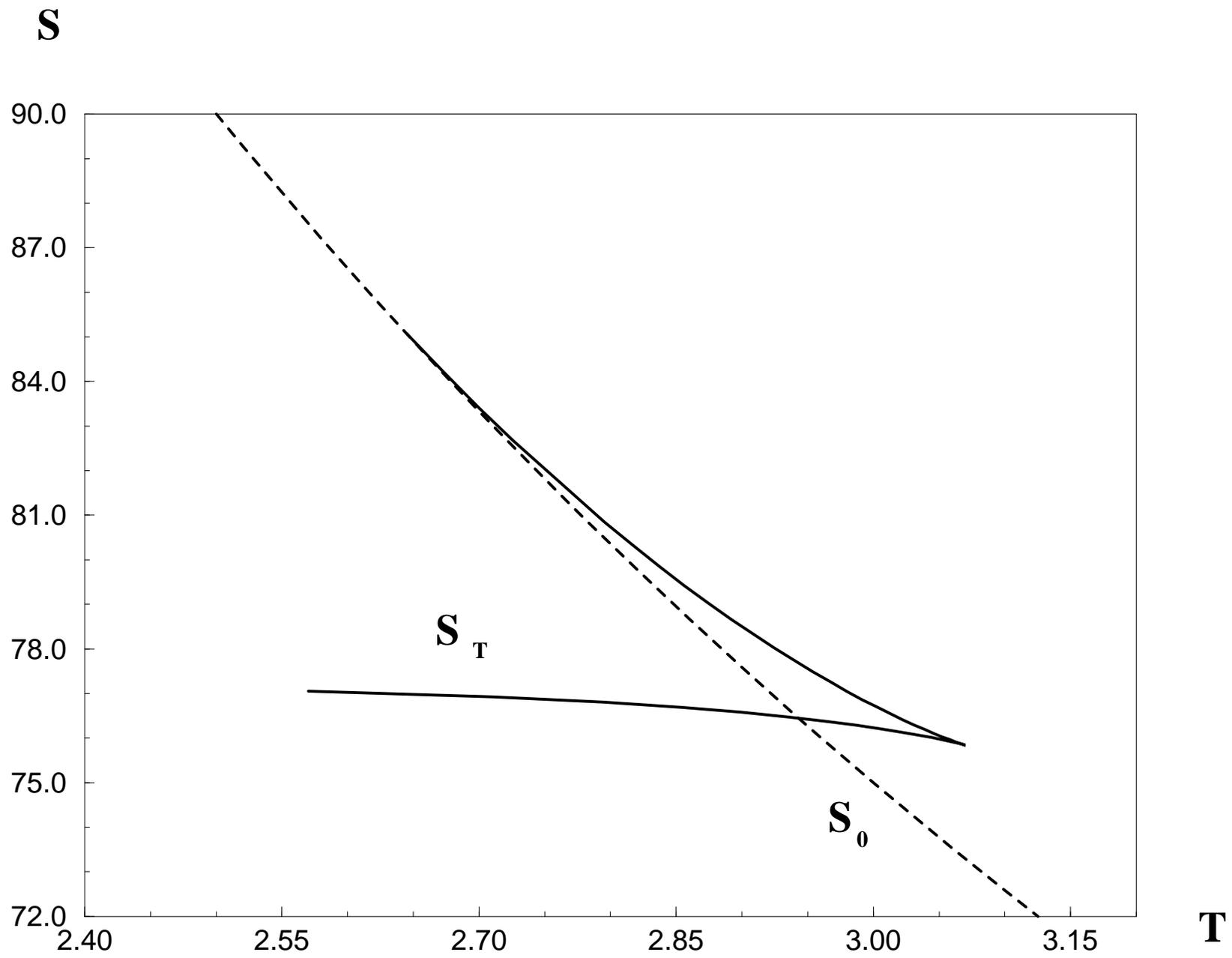
Fig. 7: The action as a function of temperature showing the first-order transition in the case with external field.\\
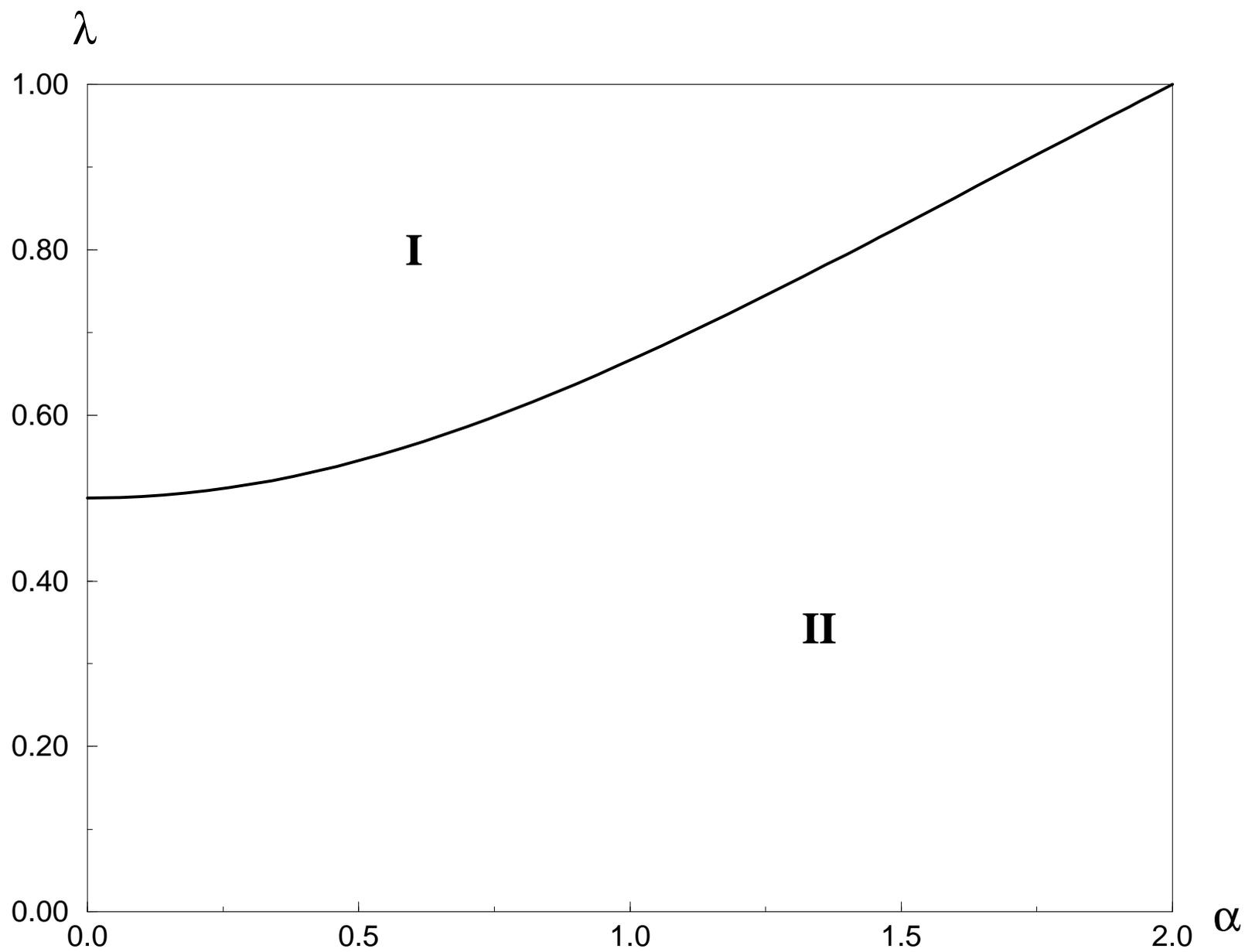
Fig. 8: The phase diagram for the orders of phase transitions in the ($\lambda,\alpha$) plane. Region I: the first-order domain, region II: the second-order domain.


\begin{thebibliography}{99}
\bibitem{Affleck} I. Affleck, Phys. Rev. Lett. {\bf 46}, 388(1981).
\bibitem{Larkin} A. I. Larkin and Yu. N. Ovchinnikov, Sov-Phys. JETP {\bf 37}, 382(1983).
\bibitem{Chudnovsky1} E. M. Chudnovsky, Phys. Rev. A {\bf 46}, 8011(1992).
\bibitem{Chudnovsky2} E. M. Chudnovsky and D. A. Garanin, Phys. Rev. Lett. {\bf 79}, 4469(1997).
\bibitem{Garanin1} D. A. Garanin and E. M. Chudnovsky, Phys. Rev. B {\bf 56}, 11102(1997); D. A. Garanin, X. Martines and E. M. Chudnovsky, Phys. Rev. B {\bf 57}, 13639(1998).
\bibitem{Liang1} J. Q. Liang, H. J. W. M\"{u}ller-Kirsten, D. K. Park, and F. Zimmerschiel, Phys. Rev. Lett. {\bf 81}, 216(1998).
\bibitem{Lee1} S.-Y. Lee, H. J. W. M\"{u}ller-Kirsten, D. K. Park, and F. Zimmerschied, Phys. Rev. B {\bf 58}, 5554(1998).
\bibitem{Garanin2} D. A. Garanin and E. M. Chudnovsky, to appear in Phys. Rev. B, cond-mat/9808037.
\bibitem{Gorokhov} D. A. Gorokhov and G. Blatter, Phys. Rev. B {\bf 56}, 3130(1997).
\bibitem{Muller-Kirsten} H. J. W. M\"{u}ller-Kirsten, D. K. Park and J. M. S. Rana, to be published.
\bibitem{Ulyanov} V. V. Ulyanov and O. B. Zaslavskii, Phys. Rep. {\bf 216}, 179(1992).
\bibitem{Hemmen} J. L. van Hemmen, A. S\"{u}t\"{o}, Europhys. Lett. {\bf 1}, 481(1986); Physica B {\bf 141}, 37(1986).
\bibitem{Enz} M. Enz and R. Schilling, J. Phys. C. {\bf 19}, 1765(1986); ibid, {\bf 19}, L771(1986).
\bibitem{Liang2} J.-Q. Liang, Y.-B. Zhang, H. J. W. M\"{u}ller-Kirsten, Jian-Ge Zhou, F. Zimmerschied and F.-C. Pu, Phys. Rev. B {\bf 57}, 529(1998).
\bibitem{Chudnovsky3} E. M. Chudnovsky and L. Gunther, Phys. Rev. Lett. {\bf 60}, 661(1988).
\bibitem{Loss} D. Loss, D. P. DiVincenzo and G. Grinstein, Phys. Rev. Lett. {\bf 69}, 3232(1992).
\bibitem{Liang4} J. Q. Liang, H. J. W. M\"{u}ller-Kirsten and Jian-Ge Zhou, Z. Phys. B {\bf 102}, 525(1997).
\bibitem{Liang3} J. Q. Liang, H. J. W. M\"{u}ller-Kirsten, Jian-Ge Zhou, F. Zimmerschied and F.-C. Pu, Kaiserslautern Report KL-TH96/13 (unpublished)
\bibitem{Park} C. S. Park, S.-K. Yoo, D. K. Park and D.-H. Yoon, cond-mat/9807344.

\end{thebibliography}
\end{document}